\documentclass{aa}
\usepackage{epsfig}
\usepackage{amssymb}
\begin{document}
\thesaurus{08.22.3; 03.13.2}
\title{Variable Stars: which Nyquist Frequency ?}
\author{L. Eyer
\and P. Bartholdi}
\institute{Observatoire de Genève, CH-1290 Sauverny, Suisse}
\offprints{L.~Eyer}
\date{Received ? ; accepted ?}
\maketitle
\begin{abstract}
In the analysis of variable stars, the problem of sampling is
central. This article focusses on the determination of the
Nyquist frequency. It is well defined in the case of regular
sampling. However, the time series of variable stars
observations are generally unevenly sampled.
Fourier analysis using the spectral window furnishes
some clues about the equivalent Nyquist frequency in the irregular case.
Often it is pushed very high, and thus very short periods can
be detected.
A specific example is shown, drawn from MACHO databases.
\end{abstract}
\section{Introduction}
When variable stars observations are analyzed, two major
problems arise.
First, there are constraints on the sampling times.
They can't always be chosen, because of weather conditions
or other observing restrictions. Secondly, the stars, even if
they are periodic, can have an extremely wide range of periods
and behaviors.
Therefore, astronomers would like to know up to which higher
limit it makes sense to search for frequency.

In the literature, there are contradictions regarding the
Nyquist frequency for irregularly sampled data. Some authors
(Press et al. 1992, Horne et al. 1986) identify it with
$1/2\overline{\delta t}$ where $\overline{\delta t}$ is the
``average'' sampling rate.
The Nyquist frequency is also often identified with
$\nu=1/2s$, where $s$ is the smallest time interval in the
sample (Scargle 1982, Roberts et al. 1987).
Nevertheless, Press et al. (1992) and Roberts et al. (1987)
remarked that frequencies can be detected above their quoted
values.

For evenly sampled data, the Nyquist frequency is defined
as $1/2\delta t$, where $\delta t$ is the time sampling step.
We will show how this quantity can be extended to the
irregular situation. In most cases, the Nyquist frequency is much
larger than usually thought. The present work is particularly
relevant to large databases that are becoming available (MACHO,
OGLE, EROS, HIPPARCOS, etc\ldots).

\section{Nyquist frequency for irregular sampling}
Let $f(t)$ be a signal function of time $t$.
We suppose that the power spectrum of the function $f$ is not
variable in time (it should be noted that this hypothesis is
stronger than stationarity). This is the case for periodic
functions. Otherwise, the Nyquist frequency is only defined
very locally and loses its sense for a given irregular sampling.
Moreover, a spectral analysis of an unstable signal which is 
undersampled may lead to wrong conclusions.

If $f$ is sampled at time $t_{i}$ we are left with
$(f_{i},t_{i})$ $i=1,\ldots,N$, where $N$ is the total number
of measurements. In the frequency domain, Deeming
(1975, 1976) showed how the irregular discretization
affects the power spectrum $F_{N}(\nu)$~: the spectrum results
from the convolution between the real spectrum $F(\nu)$ and the
spectral window $G_{N}(\nu)$. $G_{N}(\nu)$ is the Fourier
transform of $g(t)=\sum \delta(t-t_{i})$. That is
$F_{N}(\nu) = F(\nu) \otimes G_{N}(\nu)$.

We can state the result as follows.
Let\footnote{Note that if the $t_i$ are irrational, $p$ may
not exist and should then be taken as zero.} $p$ be the largest value such that $\forall \, t_{i}$,
$\,t_{i}= t_{1} + n_{i} p \:$, where
$n_{i} \in \mathbb{N}$. $p$ is
a kind of greatest common divisor (gcd) for all $(t_{i}-t_{1})$.
The Nyquist Frequency then is:
\begin{equation}
\nu_{\mbox{\tiny Ny}}=\frac{1}{2p} \geq \frac{1}{2s}.
\end{equation}
In most practical cases $p \ll s $, therefore, even with a
strong but random under-sampling, very high frequencies can be
detected.

\section{Proof}
The key function is the spectral window:
\begin{equation}
G_{N}(\nu) = \frac{\mid \sum_{k=1}^{N} \exp(i 2\pi \nu t_{k})\mid^{2}}{N^{2}}.
\end{equation}
We have $ G_{N}(\nu) \leq 1$ and $G_{N}(\nu)=1 \: \mbox{for} \:
\nu = m/p, m \in \mathbb{N}.$
Replacing $t_{k}$ with $n_{k} p + t_1$, we notice that
this function is periodic with period $1/p$
(each individual term in the sum has this same periodicity).
Moreover as $g(t)$ is real, the function
$G_{N}(\nu)$ is symmetric around the origin.
These properties of periodicity and symmetry imply that
$G_{N}(\nu)$ is also symmetric around $(2k+1)/2p$. For $k=0$,
we have $1/2p$, which is just the Nyquist frequency.
For the power spectrum, the proof is exactly the same. 
We can think of the set ${t_{i}}$ as a regular sampling with
step $p$, where some (most) data have been dropped. The more
random the data is, the smaller $p$ will be and so the higher
$\nu_{\mbox{\tiny Ny}}$.
As in the regular case, the continuous signal $f(t)$ should not
contain any component above $\nu_{\mbox{\tiny Ny}}$. Otherwise, it will be
mirrored into $2\nu_{\mbox{\tiny Ny}}-\nu$, and, without
further assumptions,
it will not be distinguishable from the mirror image. Of course,
no frequency component filtered during the measurement process
can be recovered by the random sampling. In particular, if
$\Delta t_{\mathrm{exp}}$ is the exposure time of every
measures, then in the frequency
domain, the signal is filtered by $\sin^{2}(\pi \Delta
t_{\mathrm{exp}} \nu)/(\pi \Delta t_{\mathrm{exp}} \nu)^{2}$ and
not much information will be recovered above
$\nu_{\mathrm{max}}=1/(\kappa \Delta t_{\mathrm{exp}}), \; \kappa = \mbox{2 or 3}$.

\section{Simulation}
The following numerical experiment illustrates the theory
presented above.
We chose the times as
\begin{equation}
t_{k} = \frac{ \lfloor 1000 \cdot \mbox{rnd} \rfloor }{100},
k=1,\ldots, 20
\end{equation}
where rnd are random numbers uniformly distributed on [0,1[.
Then, we built up a signal as follows
$f(t_{i})=3 \sin(2\pi \nu_{0} t_{i})$, with
$\nu_{0}=43$ (period = .0233\ldots).
\begin{center}
  \begin{tabular}{|lr||lr|} \hline
       Time &   Signal& Time & Signal  \\ \hline
       0.00  &   -2.69 & 5.59  &   0.88\\
       0.52  &    0.70 & 6.12  &  -2.55\\
       1.07  &   -2.77 & 7.27  &   2.91\\
       1.33  &   -2.21 & 7.84  &  -2.86\\
       1.77  &   -2.91 & 8.13  &   2.98\\
       2.00  &   -2.69 & 8.32  &   1.14\\
       2.77  &   -2.91 & 8.49  &  -2.99\\
       3.31  &    0.14 & 8.54  &  -1.80\\
       4.30  &   -1.40 & 9.00  &  -2.69\\
       4.49  &   -2.99 & 9.29  &   2.39\\  \hline
  \end{tabular}          
\end{center}
The smallest time interval is $s=0.05$, so $\nu_{0}$ is well
above $1/2s$, but $p$ is 0.01.
As expected, we observe that the Nyquist
frequency is at 50, and that there is no strong aliasing peak
before $\nu = 100$ (figure 1).

If we had rounded the times $t_k$ to $0.001$ in the above
example and had $p$ at $0.001$, we would have found
$\nu_{\mbox{\tiny Ny}}$ at 500. 

\begin{figure}
        \epsfxsize=8cm
        \centering{\mbox{\epsfbox{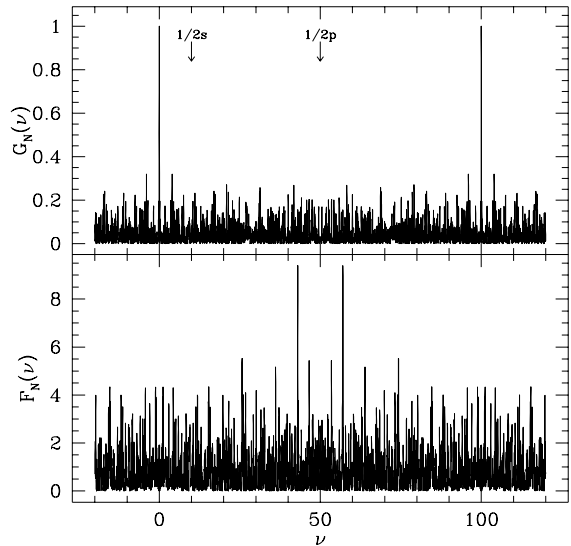}}}
        \caption{Spectral window $G_N$ and power spectrum $F_N$
        for the simulation described in the text. Both
        functions are clearly symmetrical around $\nu_{\mbox{\tiny
        Ny}}=1/2p=50$. No peak appears for $G_{N}(\nu)$ at
        $\nu=1/s$.}
\end{figure}

\section{Determination of $p$}
Theoretically, $p$ could be found using an extension to more
than 2 numbers of the Euclid gcd algorithm. But in practice,
the ${t_{k}}$ are not known to infinite precision, and the
Euclid algorithm is numerically unstable for non integer
numbers. A practical solution could then be to stop the algorithm
when the smallest number is smaller than some~$\epsilon$.
Otherwise, the spectral window is probably the most efficient
way to find $p$ in this fuzzy environment. $p$ is the inverse
of the smallest $\nu$ for which $G_{N}(\nu)$ is greater than some
value just below one.

\section{Application}
Actually, most data are obtained at approximately equal time
intervals. The fact that there is some randomness may raise
substantially the Nyquist frequency.
As a conspicuous example, we can cite Minniti et al. (1998) who
are able to find periods for $\delta$~Scuti stars with MACHO
observations of bulge fields. Indeed, the data are
separated by about one day and the periods found are of
few hours. We give an example for the $\delta$~Scuti
star 162.25348.3066. The histogram of the time intervals is
given in figure~2, its Fourier transform is shown in figure~3,
and finally the folded curve is presented in figure~4, it
leaves no doubts that the period is the right one. The
detection limit is dominated by the exposure time (150
seconds), that is about $\nu_{\mathrm{max}}=192$ $\mbox{days}^{-1}$ if
$\kappa=3$ ($P_{\mathrm{min}}=7.5 \; \mbox{min}$).

\begin{figure}
        \epsfxsize=8cm
        \centering{\mbox{\epsfbox{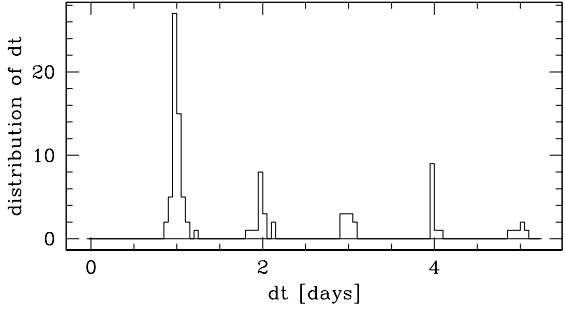}}}
	\vspace{-2mm}
        \caption{Histogram of the time intervals. The
        smallest is around 0.8 days. In total, 122 measurements
        were available.}
\end{figure}
\begin{figure}
        \epsfxsize=8cm
        \centering{\mbox{\epsfbox{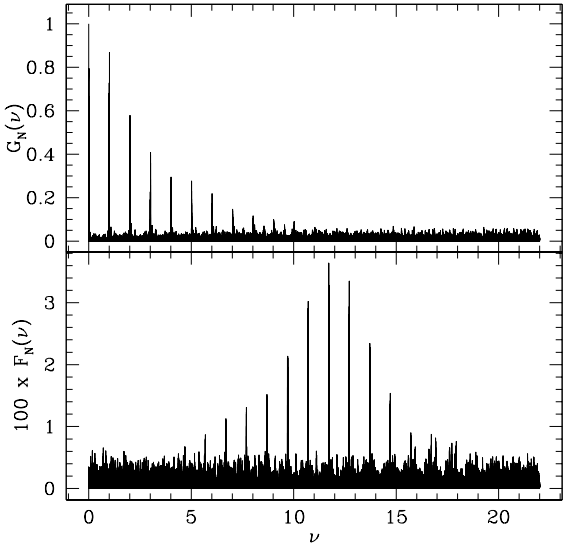}}}
	\vspace{-2mm}
        \caption{Spectral window $G_N$ and power spectrum $F_N$
        of the $\delta$~Scuti star 162.25348.3066 from MACHO.
        Although there are rather strong aliasing peaks, the
        frequency of $1/0.0854259\; \mbox{days}^{-1}$ is unambiguous.
	}
\end{figure}\begin{figure}
        \epsfxsize=8cm
        \centering{\mbox{\epsfbox{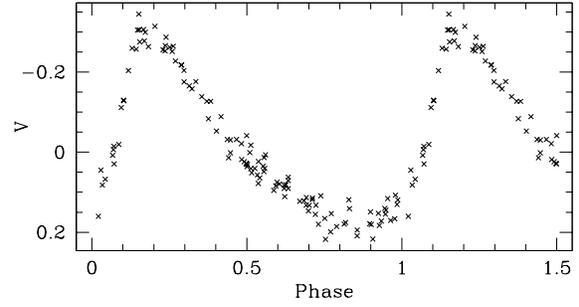}}}
        \caption{The folded curve of the $\delta$~Scuti star
        162.25348.3066 from MACHO (period $=$ 0.0854259 days).}
\end{figure}

With the HIPPARCOS mission the smallest time interval is 20
minutes. A search for shorter periods was undertaken (Eyer,
1998). Unfortunately no results came out. We can invoke that
often the spectral window is nearly symmetric around 1/40
$\mbox{min}^{-1}$ (this is due to the fact that observations
are often found in sequences, inducing strong aliasing peaks),
or that short periods are rarely stable over three years
(either period change or phase shift can occur).

\section{Reducing the amplitude of peaks in $G_N(\nu)$}
The spectral window also helps to decide when new measurements
should be made to reduce the annoying spikes in the spectral
window. We make use of the simulated data in section 4 as an
example.
A practical procedure could be to determine the polar coordinates
$\rho,\theta$ (point (1) in figure~5, where we used as a
concrete example the simulated data of section 4) of the
complex number $(1/N)\sum_{k=1}^{N} \exp(i 2\pi \nu_{h} t_{k})$
for the annoying frequency $\nu_{h}$. Point (1) is the
center of gravity of the observing points. The value of
$G_{N}(\nu_{h})$ is the distance of point (1) to the origin.
Then, the new measurement should be taken at
\begin{equation}
t_{N+1}= \left(k + \frac{\pi + \theta}{2 \pi}\right) \frac{1}{\nu_{h}}
\end{equation}
for some integer value $k$ compatible with the other
constraints.
This produces the largest reduction of the peak.
\begin{figure}
        \epsfxsize=6cm
        \centering{\mbox{\epsfbox{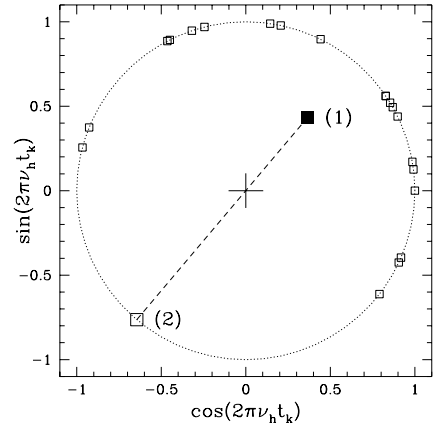}}}
        \caption{The points $\exp(i 2\pi \nu_h t_k)$ are
        displayed in the complex plane (small open squares) for
        the frequency $\nu_{h}=3.99$ corresponding
        to the highest peak of $G_{N}(\nu)$ ($ 1 < \nu < 50$,
        in figure 1).
        The black square (1) represents the center of gravity of
        these points, while the open square (2) opposed to (1)
        on the circle, is the best position for $t_{N+1}$
        (modulo $2\pi \nu_{h}$) to reduce the amplitude of
        $G_N$ at $\nu_h$.}
\end{figure}
\vspace{-3.9mm}
\acknowledgements{We would like to thank S.~Paltani, G.~Burki,
F.~Kienzle, C.~Fluetsch, D.~Kurtz and the reviewer A.~Milsztajn
for their interesting
discussions and comments. Furthermore, we thank very warmly
D.~Minniti for his helpful and efficient collaboration.}


\vspace{-2mm}
\begin{thebibliography}{}
\bibitem[1975]{Deeming:1975}
Deeming, T.J., 1975, Astrophys. Space Science 36, 137
\bibitem[1992]{Deeming:1976}
Deeming, T.J., 1976, Astrophys. Space Science 42, 257
\bibitem[1986]{Eyer:1998}
Eyer, L., 1998, PhD Thesis, Geneva University, Switzerland
\bibitem[1986]{Horne:1986}
Horne, J.H., Baliunas, S.L., 1986, ApJ 302, 757
\bibitem[1998]{Minniti:1998}
Minniti, D., Alcock, C., Alves, D.R., Axelrod, T.S., Becker,
A.C., Bennett, D.P., Cook, K.H., Freeman, K.C., Griest, K.,
Lehner, M.J., Marshall, S.L., Peterson, B.A., Quinn, P.J.,
Pratt, M.R., Rodgers, A.W., Stubbs, C.W., Sutherland, W.,
Tomaney, A., Vandehei, T., Welch, D., 1998, IAU Symp. 189,
p.~293
\bibitem[1992]{Press:1992}
Press, W.H., Teukolsky, S.A., Vetterling, W.T., Flannery, B.P.,
1992, Numerical Recipes in Fortran, Cambridge University Press
\bibitem[1987]{Roberts:1987}
Roberts, D.H., Leh\'ar, J., Dreher, J.W., 1987, AJ 90, 968
\bibitem[1982]{Scargle:1982}
Scargle, J.D., 1982, ApJ 263, 835
\end{thebibliography}
\end{document}